\documentclass[aps]{revtex4-1}
\usepackage{amsmath}
\usepackage{amsfonts}
\usepackage{graphicx}
\usepackage{listings}
\usepackage{color}

\newcommand{\done}{D^{\mathrm{(1)}}}
\newcommand{\donetilde}{\tilde{D}^{\mathrm{(1)}}}
\newcommand{\donehat}{\hat{D}^{\mathrm{(1)}}}
\newcommand{\dtwo}{D^{\mathrm{(2)}}}
\newcommand{\dtwotilde}{\tilde{D}^{\mathrm{(2)}}}
\newcommand{\dtwohat}{\hat{D}^{\mathrm{(2)}}}
\newcommand{\dnhat}{\hat{D}^{\mathrm{(}n\mathrm{)}}}
\newcommand{\rw}{R_{\mathrm{W}}}
\newcommand{\rmd}{\mathrm{d}}
\newcommand{\rme}{\mathrm{e}}

\begin{document}
\title{Estimation of drift and diffusion functions from time series
  data: A maximum likelihood framework}

\author{David Kleinhans} \email{david.kleinhans@gu.se}
\affiliation{University of Gothenburg, Department of Biological and
  Environmental Sciences, Box 461, SE-405 30 G\"{o}teborg, Sweden}%

\begin{abstract}
  Complex systems are characterized by a huge number of degrees of
  freedom often interacting in a non-linear manner. In many cases
  macroscopic states, however, can be characterized by a small number
  of order parameters that obey stochastic dynamics in time.  Recently
  techniques for the estimation of the corresponding stochastic
  differential equations from measured data have been introduced. This
  contribution develops a framework for the estimation of the
  functions and their respective (Bayesian posterior) confidence
  regions based on likelihood estimators. In succession approximations
  are introduced that significantly improve the efficiency of the
  estimation procedure.  While being consistent with standard
  approaches to the problem this contribution solves important
  problems concerning the applicability and the accuracy of estimated
  parameters.
\end{abstract}


\maketitle

\section{\label{sec:introduction}Introduction}
Complex systems are characterized by a huge number of degrees of
freedom. In presence of nonlinear interactions such systems can form
macroscopically ordered states characterized by a rather small number
of order parameters that dominate the dynamics at large time scale
\cite{Haken:Synergetics}. The microscopic degrees of freedom force the
system stochastically at small time scales which can have a
significant impact on the dynamics, in particular far from
equilibrium. A natural framework for modelling the dynamics of complex
systems are stochastic differential equations of the form\footnote{For
  reasons of clarity we here restrict to a one dimensional
  process. Generalisation to higher dimensions, however, is
  straightforward (see e.g.\
  \cite{Friedrich08:Ubersicht,Risken,Gardiner}).}
\begin{equation}
  \label{eq:langevin}
  \frac{\rmd}{\rmd t}x(t)=\done(x,t)+\sqrt{\dtwo(x,t)}\Gamma(t)
\end{equation}
characterized by the drift and diffusion functions $\done(x,t)$ and
$\dtwo(x,t)$. Here $\Gamma(t)$ is a Gaussian white noise source with
correlation function
$\left\langle\Gamma(t)\Gamma(t')\right\rangle=2\delta(t-t')$. The term
$\sqrt{\dtwo(x,t)}\Gamma(t)$ is interpreted according to It\^o's
definition of stochastic integro-differential equations
\cite{Risken}. Drift and diffusion account for the deterministic and
the stochastic contributions respectively and are directly linked to
the deterministic dynamics of order parameters and the character of
the intrinsic dynamical noise. Often these functions cannot be derived
explicitly since the interactions of the microscopic degrees of
freedom are unknown.

As recently demonstrated drift and diffusion $\done(x,t)$ and
$\dtwo(x,t)$ can directly be estimated from ensembles of time series
$x(t)$ measured on the respective systems by the definition
\cite{Siegert98,Friedrich00PhysA}
\begin{equation}
  \label{eq:direst-est} 
  \dnhat(x',t):=\lim\limits_{\tau\to 0}\frac{1}{n!\tau}\left\langle
    \left(x(t+\tau)-x(t)\right)^n|x(t)=x'\right\rangle\:,\:n=1,2\quad.
\end{equation}
Here $x(t)=x'$ indicates, that only increments close to $x'$ are
considered for averaging, where the interpretation of \emph{close}
depends on the kernel or the binning used for estimation
\cite{Lamouroux09}.  If the process can assumed to be stationary,
i.e.\ $\done$ and $\dtwo$ do not depend on time explicitly, ensemble
averages correspond to time averages and these expressions can be
calculated from a single time series.  A qualitative estimation of
drift and diffusion functions from (\ref{eq:direst-est}) e.g.\ by
evaluation of the expression for the smallest available time increment
$\tau$ is straightforward and has very low computational demands. This
approach in the following will be termed \emph{direct} estimation. It
successfully has been applied to different systems ranging from
turbulent flows \cite{Friedrich97,Luck99} and wind power converters
\cite{Gottschall07} to financial markets \cite{Friedrich00PRL},
traffic flow data \cite{Kriso02}, and medical applications
\cite{Kuusela04,Prusseit07}. For a recent comprehensive review both on
the estimation procedure and on applications the reader is referred to
\cite{Friedrich11}.

If quantitative reliable estimates of drift and diffusion are required
the explanatory power of the \emph{direct} estimation
(\ref{eq:direst-est}) is limited, since it is sensitive to the
sampling frequency, the estimation kernel used, the amount of data
used for estimation, and systematic deviations e.g.\ due to
measurement noise. In the course of time many of these drawbacks have
been addressed and solved, e.g.\ through corrections of the original
estimation procedure for finite sampling frequencies
\cite{Sura02,Ragwitz01,FriedrichComment02,Gottschall08NJP}, through
new approaches explicitly circumventing the limit of high sampling
frequencies \cite{Kleinhans05,Lade09,Honisch11}, and through
explicitly considering the presence of measurement noise in the
estimation procedure \cite{Boettcher06,Lind10,Lehle10}. As a matter of
fact many of these advancements, however, significantly complicate the
estimation.

This work exhibits an alternative approach to the topic, since it
focuses on joint probability distributions of observed data and aims
to find the model underlying with maximum likelihood. Likelihood
estimators for the estimation of drift and diffusion functions already
were investigated intensively in the past decade, in particular in the
mathematics community and with respect to financial applications
\cite{Sahalia02,Lo88,Durham02}. Recently, a couple a contributions
emphasized their applicability in a more general context
\cite{Wu11,Ohkubo11,Hurn03}.  One particular advantage of the maximum
likelihood approach, however, not sufficiently discussed so far is the
fact that the results easily can be interpreted in a quantitative
manner since confidence regions for parameters can be calculated for
any parameter estimated.  The aims of this work are twofold. At first
I give an comprehensive introduction into the application of maximum
likelihood estimators. Under certain circumstances that will be
discussed sequentially the estimation procedure drastically can be
reduced. Methods straightforwardly applicable in the respective cases
are discussed in detail; examples for their numerical implementation
are included in the appendix. For each of the procedures described, at
second, methods for evaluating the accuracy of the estimates are
described.

We in section \ref{sec:estim-param-drift} exemplary start from a
rather simple example, an Ornstein-Uhlenbeck process with known finite
time propagator. On this example the concepts of maximum likelihood
estimation of parameters and of Bayesian posterior confidence
intervals for the individual parameters (section
\ref{sec:bayes-post-conf}) are introduced.  In principle this
procedure also is applicable to the estimation of general parametrized
drift and diffusion functions \cite{Kleinhans05,Kleinhans07MLE},
though at rather high computational costs. The effort, however,
reduces significantly for measurements available at high frequencies,
as discussed in sections
\ref{sec:estim-gener-drift}-\ref{sec:effic-appr-estim}.

At first in section \ref{sec:estim-gener-drift} an approximation to
the small-time propagator is implemented in the estimation procedure,
which makes it applicable to more complex problems at reasonable
effort. It is shown, that the estimation results are satisfactory even
at small but finite time lags. If drift and diffusion in addition are
assumed to be smooth functions this approach can be extended to a
procedure for the non-parametric estimation of drift and diffusion, as
outlined in section \ref{sec:non-param-estim}.  The results are
demonstrated to be consistent with the \emph{direct} estimation
procedure, (\ref{eq:direst-est}). In section
\ref{sec:effic-appr-estim} this procedure is developed further and
applied to the optimization of parametric functions for drift and
diffusion, which results in a tremendous gain in numerical efficiency
with respect to previous approaches. The method developed in this
section exhibits the main contribution of this work. It impresses
through its accuracy even at moderate amount of data and since its
implementation neither involves numerical integration nor other
computationally demanding operations it is straightforward to be
applied even to large data sets.

This work aims to introduce a new approach to the estimation of drift
and diffusion and to discuss it in detail. In order to be able to
focus on the problems intrinsic to the estimation procedure we work on
synthetic data generated numerically from a stochastic differential
equation of the form (\ref{eq:langevin}). That is, we at this stage
explicitly exclude problems arising from models not conform with the
data e.g.\ due to measurement noise. The impact of such effects needs
to be investigated separately in future.

\section{\label{sec:estim-param-drift}Estimation of parametric drift
  and diffusion functions at arbitrary time increment}
This section aims to introduce the basic methodology of maximum
likelihood estimation of parametric drift and diffusion functions.

Let us assume that we have measurements $x_0(t_0), x_1(t_1),\ldots
x_N(t_N)$ on a Markovian system at times $t_i=t_0+i\tau$ and let us
assume that the dynamics is stationary in time. If we furthermore have
a model for the propagator at the time interval $\tau$ in form of the
conditional probability density function
$p(x_{i}|x_{i-1};\boldsymbol{\Omega},\tau)$, that is consistent with
the data set and that depends on set of parameters
$\boldsymbol{\Omega}\in\mathcal{P}\subseteq \mathbb{R}^m$: How should
we determine the unknown parameters $\boldsymbol{\Omega}$, that best
match the measurements?

A natural approach to this problem is to consider the probability of a
certain $\boldsymbol{\Omega}$ given the realized measurements. In
general this probability can be calculated using Bayes theorem,
\begin{equation}
  \label{eq:bayes}
  p_{\boldsymbol{\Omega}}(\boldsymbol{\Omega}|x_N,\ldots,x_0;\tau)= \frac{p_{\boldsymbol{x}}(x_N,\ldots,x_1|x_0;\boldsymbol{\Omega},\tau)f_{\boldsymbol{\Omega}}(\boldsymbol{\Omega})}{f_{\boldsymbol{x}}(x_N,\ldots,x_1|x_0;\tau)}\quad.
\end{equation}
Here, $f_{\boldsymbol{\Omega}}$ and $f_{\boldsymbol{x}}$ are prior
distributions of $\boldsymbol{\Omega}$ and the measurements
$\boldsymbol{x}$. For reasons of clarity we from now on drop the
subscripts of probability density functions if unambiguously defined
through their arguments. If we do not have particular information on
the prior distribution it is reasonable to assume that priors are
uniformly distributed on appropriate intervals. Then (\ref{eq:bayes})
yields
\begin{equation}
  \label{eq:bayes_uniform}
  p(\boldsymbol{\Omega}|x_N,\ldots,x_0;\tau)\sim p(x_N,\ldots,x_1|x_0;\boldsymbol{\Omega},\tau)\quad,
\end{equation}
implying, that the probability of a certain set of parameters
$\boldsymbol{\Omega}$ is proportional to the joint probability of the
realisation of the measurement data under the respective model. The
idea of the maximum likelihood approach is to determine a set of
parameters $\boldsymbol{\tilde{\Omega}}\in\mathcal{P}$ maximising
expression (\ref{eq:bayes_uniform}).

Since the logarithm is a strictly monotonous function maximising the
likelihood of $\boldsymbol{\Omega}$ is equivalent to maximizing
$\log\left[ p(\boldsymbol{\Omega}|x_N,\ldots,x_0)\right]$. Since we
have a Markov process the conditional pdf
$p(x_N,\ldots,x_1|x_0;\boldsymbol{\Omega})$ degenerates and we can
define
\begin{equation}
  \label{eq:log_likelihood_function}
  L(\boldsymbol{\Omega}):=\sum\limits_{i=1}^N\log\left[p(x_{i}|x_{i-1};\boldsymbol{\Omega},\tau)\right]\quad,
\end{equation}
which we call the log-likelihood function and which we aim to maximize
with respect to $\boldsymbol{\Omega}$.

One model, on which the optimisation process can be demonstrated
exemplarily and which will act as a baseline case in this manuscript
is the Ornstein-Uhlenbeck process with linear repulsive drift
$\done(x)=-\gamma x$ and constant diffusion $\dtwo(x)=Q$. For the
Ornstein-Uhlenbeck process the explicit form of the finite time
propagator is
\begin{equation}
  \label{eq:ou-finite-time-prop}
  p(x_{i+1}|x_{i};(\gamma,Q),\tau)=\sqrt{\frac{\gamma}{2\pi
      Q(1-\rme^{-2\gamma\tau})}}\exp\left(-\frac{\gamma\left[x_{i+1}-x_{i}\rme^{-\gamma\tau}\right]^2}{2Q(1-\rme^{-2\gamma\tau})}\right)\quad.
\end{equation}
In general the parameters $(\tilde{\gamma},\tilde{Q})$ with maximum
likelihood can now be obtained through numerical optimization of the
log-likelihood function (\ref{eq:log_likelihood_function}), which is
quite efficient in cases where the propagator is known
analytically. For the Ornstein-Uhlenbeck process the maximum can be
derived analytically. The parameters maximizing the likelihood
function only depend on certain statistical properties of the measured
sample,
\begin{equation}
  \label{eq:op-opt}
  \tilde{\gamma}=\frac{1}{\tau}\log\left[\frac{\left\langle
        x_i^2\right\rangle_i}{\left\langle
        x_ix_{i-1}\right\rangle_i}\right]\quad\mbox{and}\quad \tilde{Q}=\frac{\tilde{\gamma}\left\langle\left(x_i-x_{i-1}\rme^{-\tilde{\gamma}\tau}\right)^2\right\rangle_i}{1-\rme^{-2\tilde{\gamma}\tau}}\quad.
\end{equation}
Table \ref{tab:ou-results} shows that $(\tilde{\gamma},\tilde{Q})$
converges to the correct $(\gamma,Q)$ for $N\to\infty$ as the
statistics converge with increasing sample size.

\begin{table*}
  \begin{tabular}{r@{\hspace{1em}}rl@{\hspace{1em}}rl}
    \hline\hline
    \multicolumn{1}{c}{$N$}&\multicolumn{2}{c}{$\tilde{\gamma}$
      ($\mathcal{C}_\gamma$)$_{\nu_\gamma}$}&\multicolumn{2}{c}{$\tilde{Q}$
      ($\mathcal{C}_Q$)$_{\nu_Q}$}\\\hline
    $ 10 $&$ -1.700 $&$( -43.784 , 40.289 )_{ 0.949 }$&$ 0.509 $&$( 0.228 , 1.516 )_{ 0.878 }$\\
    $ 100 $&$ 2.734 $&$( -1.874 , 7.344 )_{ 0.950 }$&$ 0.974 $&$( 0.746 , 1.305 )_{ 0.945 }$\\
    $ 1000 $&$ 1.444 $&$( 0.357 , 2.530 )_{ 0.950 }$&$ 0.955 $&$( 0.875 , 1.043 )_{ 0.950 }$\\
    $ 10000 $&$ 1.193 $&$( 0.889 , 1.497 )_{ 0.950 }$&$ 1.007 $&$( 0.980 , 1.035 )_{ 0.946 }$\\
    $ 100000 $&$ 1.051 $&$( 0.960 , 1.141 )_{ 0.952 }$&$ 1.001 $&$( 0.992 , 1.010 )_{ 0.951 }$\\
    $ 1000000 $&$ 1.000 $&$( 0.972 , 1.028 )_{ 0.955 }$&$ 1.000 $&$( 0.997 , 1.003 )_{ 0.962 }$\\
    \hline\hline
  \end{tabular}
  \caption{\label{tab:ou-results}Parameters $\tilde{\gamma}$ and
    $\tilde{Q}$ estimated from realisations of an  Ornstein-Uhlenbeck
    by evaluation of  
    (\ref{eq:op-opt}) for different sample sizes $N$. The brackets
    indicate Bayesian posterior
    confidence intervals (BpCIs) based on the log-likelihood offset
    $\rw(0.95)$ obtained from Wilks' theorem, (\ref{eq:wilks-one-d}). Subscripts to the
    brackets mark the realized confidence levels as calculated from
    equation (\ref{eq:conf-level-nu-of-R-i}). The data set was generated
    numerically by sampling (\ref{eq:ou-finite-time-prop})
    with $\gamma=1$, $Q=1$ and $\tau=0.01$. With increasing
    $N$ the estimate $(\tilde{\gamma},\tilde{Q})$ converges to these parameter
    values. BpCIs accurately seem to model the
    statistical uncertainties, since the true values used for simulation
    always are within the confidence interval. The numerically calculated
    confidence levels in all but one cases are very close to the desired
    value of $0.95$ suggesting that  the log-likelihood offset calculated from
    Wilks' theorem is accurate enough for the present purpose.}
\end{table*}

\section{\label{sec:bayes-post-conf}Bayesian posterior confidence
  intervals (BpCIs) and Wilks' theorem}

With the maximum likelihood approach it is straightforward to
investigate the accuracy of the estimated parameters. For this purpose
we define Bayesian posterior confidence regions (BpCR)
$\mathcal{C}\in\mathcal{P}$ for the estimates as compact sets around
the optimal parameters $\boldsymbol{\tilde{\Omega}}$ that contain the
true parameter with a certain probability. In practise the confidence
level $\nu$ often is set to $\nu=95\%$, which also is used throughout
this work.\\
BpCRs differ subtly from classical confidence regions as typically
defined in statistics (cCR): They define sets which contain the true
parameter with a certain probability $\nu$ given that the priors were
chosen appropriately and that the data is consistent with the general
model. One could argue that this is the best one can do from a single
measurement on a system. In contrast cCRs consists of all parameters
for that the observed realisation would belong to the $\nu100\%$ most
probable realisations. The differences between these definitions and
their respective advantages intensively have been debated since a long
time already (see e.g.\ \cite{Jaynes76}) and we are not intending to
join this debate here. The most important point perhaps is, that the
deviations between the definitions are rather small if the number of
observations is sufficiently large \cite{Jaynes76}, as we also will
see on a practical example in section \ref{sec:non-param-estim}. And
that BpCIs from the technical point of view have significant
advantages, since they easily can be determined numerically.

The boundaries of BpCRs still allow for some freedom. The standard
definition requires to choose the boundaries corresponding to the
smallest BpCRs \cite{Jaynes76}. A technically more convenient
alternative in connection with likelihood estimators is to consider a
threshold $r$ of the likelihood ratio ${\exp
  [L({\boldsymbol{\Omega}})]}/{\exp[L(\boldsymbol{\tilde{\Omega}})]}$,
which is bound from above by $1$ at
$\boldsymbol{\Omega}=\boldsymbol{\tilde{\Omega}}$. The BpCRs
associated with $r$ then are defined to extend to all parameters
$\boldsymbol{\Omega}$ with ${\exp
  [L({\boldsymbol{\Omega}})]}/{\exp[L(\boldsymbol{\tilde{\Omega}})]}\ge
r$ in the compact set containing $\boldsymbol{\tilde{\Omega}}$. This
definition of BpCRs that will be applied throughout this work.
Parameters at the boundaries of $\mathcal{C}$ have in common that they
are realized with a probability which is only a fraction $r$ of the
probability for the most likely set of parameters.  Since it often is
advantageous to work with log-likelihoods, a more practicable form of
the set $\mathcal{C}$ is
\begin{equation}
  \label{eq:likelihood-ratio}
  \mathcal{C}=\left\{\boldsymbol{\Omega}\in\mathcal{P}|L({\boldsymbol{\Omega}})\ge L(\boldsymbol{\tilde{\Omega}})+R\right\}
\end{equation}
with $R:=\log(r)$. The boundaries of $\mathcal{C}$ can be iterated numerically by root finding algorithms.\\
So far $R$ was not directly related to a certain confidence level. If
the integral $\int_{\boldsymbol{\Omega}\in\mathcal{P}}\rmd^m\Omega'\
\exp[L(\boldsymbol{\Omega'})]$ exists the confidence level, however,
is a function of $R$ that can be calculated as
\begin{equation}
  \label{eq:conf-level-nu-of-R}
  \nu(R)=\frac{\int_{\boldsymbol{\Omega}\in\mathcal{C}}\rmd^m\Omega'\
    \exp[L(\boldsymbol{\Omega'})]}{\int_{\boldsymbol{\Omega}\in\mathcal{P}}\rmd^m\Omega'\
    \exp[L(\boldsymbol{\Omega'})]}\quad.  
\end{equation}
In principle $R$ now could be iterated until the desired confidence
level for the maximum likelihood estimate is achieved. The
corresponding region $\mathcal{C}$ then with probability $\nu$ would
contain the correct estimate of the parameters.

Instead of considering sets specifying entire regions in parameter
space typically confidence intervals in the individual components of
$\boldsymbol{\tilde{\Omega}}$ are derived independently, since these
errors are much easier to discuss.  We again stick to the Bayesian
framework and now define Bayesian posterior confidence
\emph{intervals} (BpCIs).  Formally these intervals can be obtained by
restricting the evaluation of equations (\ref{eq:likelihood-ratio})
and (\ref{eq:conf-level-nu-of-R}) to a certain subset of $\mathcal{P}$
only. For component $i$ of $\boldsymbol{\Omega}$,
$\Omega_i\in\mathcal{P}_i$, this corresponds to the respective
expressions
\begin{subequations}
  \label{eq:confidence-formulas-i}
  \begin{eqnarray}
    \label{eq:likelihood-ratio-i}
    \mathcal{C}_i(R_i)&=&\left\{\Omega_i\in\mathcal{P}_i|L(\tilde{\Omega}_1,\ldots,\tilde{\Omega}_{i-1},\Omega_i,\tilde{\Omega}_{i+1},\ldots,\tilde{\Omega}_{m})\ge
      L(\boldsymbol{\tilde{\Omega}})+R_i\right\}\quad,\\ 
    \label{eq:conf-level-nu-of-R-i}
    \nu_i(R_i)&=&\frac{\int_{\Omega_i\in\mathcal{C}_i(R_i)}\rmd\Omega_i'\
      \exp[L(\tilde{\Omega}_1,\ldots,\tilde{\Omega}_{i-1},\Omega_i',\tilde{\Omega}_{i+1},\ldots,\tilde{\Omega}_{m})]}{\int_{\Omega_i\in\mathcal{P}_i}\rmd\Omega_i'\
      \exp[L(\tilde{\Omega}_1,\ldots,\tilde{\Omega}_{i-1},\Omega_i',\tilde{\Omega}_{i+1},\ldots,\tilde{\Omega}_{m})]}\quad.  
  \end{eqnarray}
\end{subequations}
Since we are now working in one dimension only the numerical
evaluation of the expressions is straightforward. An alternative to
iteration of equations (\ref{eq:confidence-formulas-i}) for the
determination of the desired $R_i$ is the application of Wilks'
theorem, which explicitly provides $R(\nu)$ in the asymptotic limit of
large test statistics \cite{Wilks38}. For one degree of freedom (which
is the single component under consideration in equations
(\ref{eq:confidence-formulas-i})) the asymptotic results are
\begin{equation}
  \label{eq:wilks-one-d}
  \rw(\nu):=-\frac{1}{2}F^{-1}_{\chi^2}(\nu,1)\quad,
\end{equation}
where $F^{-1}_{\chi^2}(\nu,1)$ is the inverse cumulative distribution
function of the $\chi^2$ distribution with one degree of freedom
evaluated at $\nu$. For the $95\%$-significance level one obtains
$\rw(0.95)=-1.92073$.

For the Ornstein-Uhlenbeck process investigated in section
\ref{sec:estim-param-drift} BpCIs for the parameters $\gamma$ and $Q$
were calculated by numerical evaluation of the interval boundaries in
(\ref{eq:likelihood-ratio-i}) for the log-likelihood offset
$\rw(0.95)$. For verification of the confidence level subsequently
(\ref{eq:conf-level-nu-of-R-i}) was evaluated. The results are listed
in table \ref{tab:ou-results} as a function of sample size. Obviously
BpCIs for the Ornstein-Uhlenbeck process accurately reflect the
statistical uncertainties even at small sample sizes. The intervals in
all cases include the true parameters used for numerical generation of
the sample. The accuracy of Wilks' theorem in this example is
satisfactory for samples of size $N=100$ and larger. For this reason
the computationally rather expensive iteration of equations
(\ref{eq:confidence-formulas-i}) might not be needed in many cases of
practical relevance. For small samples validation of the confidence
level with (\ref{eq:conf-level-nu-of-R-i}), however, is recommended if
accuracy is crucial.

From the example discussed in section \ref{sec:estim-param-drift} it
became evident, that calculation of maximum likelihood estimates is
straightforward if the propagator is known explicitly. In addition
BpCRs or BpCIs can be iterated numerically at a desired confidence
level, which exhibits the main advantage of this approach.

In principal this procedure is also applicable to the estimation of
general drift and diffusion functions, since the propagator can be
calculated for finite times by (numerical) solution of the
Fokker-Planck equation \cite{Risken,Sahalia02,Kleinhans07MLE}. Since
calculations in this case are computationally very demanding for large
data sets data aggregation e.g.\ by approximation of the fine-grained
probability distribution function is recommended, which reduces the
number of function evaluations required and transforms the problem
into minimizing the Kullback-Leibler distance between the observed and
modelled propagators \cite{Kleinhans05,Kleinhans07MLE}. In this sense
maximum likelihood estimation is equivalent to the maximum entropy
principle introduced by Jaynes' \cite{Jaynes57}. Accurate estimates of
BpCIs and confidence levels, however, in the general case barely can
be obtained at reasonable numerical effort. For this reason the
remaining part of the manuscript is devoted to approximations
generally applicable in case of high sampling frequencies and smooth
drift and diffusion functions, which drastically improve the numerical
efficiency of the estimation process.
 
\section{\label{sec:estim-gener-drift}Estimation of parametric drift
  and diffusion functions at small time increment}
The \emph{direct} estimation procedure briefly described in section
\ref{sec:introduction} requires evaluation of moments in the limit
$\tau\to 0$, i.e.\ it requires measurement data to be available at
high sampling frequency. Also the maximum likelihood estimation
considerably can benefit from high sampling frequencies.

The main drawback of a maximum likelihood estimation of general drift
and diffusion functions is, that for the propagator in the
log-likelihood function (\ref{eq:log_likelihood_function}) is not
available analytically but from numerical solution of the
Fokker-Planck equation only. At sufficiently small time increments
this propagator, however, can be approximated by
\cite{Risken,Friedrich11,Sahalia02}
\begin{equation}
  \label{eq:propagator-small-tau}
  p(x_{i+1}|x_{i};\boldsymbol{\Omega},\tau)\approx \frac{1}{2\sqrt{\pi
      \dtwo(x_{i},\boldsymbol{\Omega})\tau}}\exp\left[-\frac{\left(x_{i+1}-x_{i}-\done(x_{i},\boldsymbol{\Omega})\tau\right)^2}{4\dtwo(x_{i},\boldsymbol{\Omega})\tau}\right]\quad.
\end{equation}
This expression directly can be used to perform a maximum likelihood
estimation with Bayesian posterior confidence intervals for the
respective parameters as introduced in great detail in the previous
section, but now for general parametrizations of drift and diffusion
functions. Since (\ref{eq:propagator-small-tau}) is only valid at
small $\tau$ systematic errors, however, are expected if data is not
available at sufficiently high sampling frequencies.

As an example for the application of the estimation procedure we come
back to an Ornstein-Uhlenbeck process. Samples of length $N=100000$
were simulated with $\done(x)=-x$ and $\dtwo(x)=1$ at different time
increments $\tau$ between $0.001$ and $1.000$. On each of the time
series the parametrized drift and diffusion functions
\begin{subequations}
  \label{eq:ddfunc-5pars}
  \begin{eqnarray}
    \label{eq:ddfunc-5pars-drift}
    \done(x,\boldsymbol{\Omega})&=&\Omega_1x+\Omega_2
    x^2+\Omega_3x^3\\
    \label{eq:ddfunc-5pars-diff}
    \dtwo(x,\boldsymbol{\Omega})&=&\Omega_4+\Omega_5 x^2
  \end{eqnarray}
\end{subequations}
were optimized with respect to $\boldsymbol{\Omega}$ by maximizing the
log-likelihood function (\ref{eq:log_likelihood_function}) with the
small-$\tau$ propagator (\ref{eq:propagator-small-tau}). BpCIs were
calculated at the $95\%$-level based on the log-likelihood offset
$\rw(0.95)$. An example of the numerical implementation of the
estimation procedure is the procedure \emph{parMLE} listed in appendix
\ref{sec:pyth-code:-impl}. Estimation results are compiled in table
\ref{tab:ou-small-tau-nobins}.

\begin{turnpage}
  \begin{table}
    \begin{tabular}{l@{\hspace{1em}}rl@{\hspace{1em}}rl@{\hspace{1em}}rl@{\hspace{1em}}rl@{\hspace{1em}}rl}
      \hline\hline
      \multicolumn{1}{c}{$\tau$}&\multicolumn{2}{c}{$\tilde{\Omega}_1$}&\multicolumn{2}{c}{$\tilde{\Omega}_2$}&\multicolumn{2}{c}{$\tilde{\Omega}_3$}&\multicolumn{2}{c}{$\tilde{\Omega}_4$}&\multicolumn{2}{c}{$\tilde{\Omega}_5$}
      \\\hline
      $0.001$&$-1.23$&$(-1.54  ,-0.93 ) $&$ -0.01$&$(-0.19 ,0.18 ) $&$ 0.02$&$(-0.07 ,0.11 ) $&$1.00 $&$( 1.01,1.02 ) $&$0.00 $&$(-0.01 ,0.00 )$\\
      $0.010$&$-0.95 $&$( -1.04, -0.86) $&$ 0.04$&$( -0.02, 0.09) $&$ -0.04$&$( -0.06, -0.01)$&$ 0.99$&$( 0.98, 1.00) $&$0.00 $&$( 0.00,0.01 )$\\
      $0.100$&$ -0.95$&$( -0.98, -0.92) $&$ 0.01$&$(-0.01 ,0.02 ) $&$ 0.00$&$(-0.01 ,0.00 )$&$0.91 $&$( 0.90, 0.92) $&$0.00 $&$(-0.01 , 0.00)$\\
      $1.000$&$-0.63 $&$(-0.63 ,-0.62 ) $&$0.00 $&$(0.00 ,0.01 ) $&$0.00 $&$(0.00 ,0.00 )$&$ 0.43$&$( 0.43, 0.44) $&$ 0.00$&$(0.00 ,0.00 )$\\
      \hline\hline
    \end{tabular}
    \caption{\label{tab:ou-small-tau-nobins}Parameters $\tilde{\Omega}_1$ to
      $\tilde{\Omega}_5$ of the drift and diffusion functions
      (\ref{eq:ddfunc-5pars}) estimated from realisations of an Ornstein-Uhlenbeck
      process  by maximization of (\ref{eq:log_likelihood_function}) with a
      small-$\tau$ approximation of the propagator,
      (\ref{eq:propagator-small-tau}).  The brackets
      indicate Bayesian posterior
      confidence intervals (BpCIs) at confidence level $\nu=0.95$, where the log-likelihood
      offset was calculated from  Wilks'
      theorem. The estimation procedure  was
      applied to samples of length $N=100000$
      generated at different time lag $\tau$ from an Ornstein-Uhlenbeck
      process with $\Omega_1=-1$, $\Omega_4=1$, and $\Omega_2=\Omega_3=\Omega_5=0$.}
  \end{table}
\end{turnpage}
 
The estimation procedure involved the optimization of five parameters
of the low order polynomials used for $\done$ and $\dtwo$. The result
of the estimation procedure was not sensitive to the starting values
(for the actual calculations the starting value $\Omega_i=1$ was used
for all $i$). At the two highest sampling rates $\tau=0.001$ and
$\tau=0.01$ the estimation procedure with the approximated propagator
produces accurate results. This demonstrates, that the small-$\tau$
approximation actually is applicable for data analysis purposes at
sufficiently large sampling frequencies. For lower sampling
frequencies (i.e.\ larger $\tau$) the estimates for $\Omega_1$ and
$\Omega_4$ deviate from the true parameters since the finite-$\tau$
approximation of the propagator is not any more valid to a sufficient
extent. These deviations are
systematic and therefore not reflected by the BpCIs.\\
The BpCIs for the parameters $2$, $3$ and $5$ always contain the value
$0$ suggesting correctly that the respective coefficients in the drift
and diffusion functions might not be required. In many applications
this is a valuable information since more appropriate estimators might
be applicable if the model can be reduced.

\section{\label{sec:non-param-estim}Estimation at small time increment
  and stepwise constant drift and diffusion: A non-parametric procedure}
In the preceding section an approximation applicable at high sampling
frequencies was introduced. The estimation becomes even more efficient
if we, in addition, approximate the resulting drift and diffusion by
piecewise constant functions. Then maximum-likelihood estimation even
is feasible for the non-parametric estimation of drift and diffusion
functions as it will be demonstrated in this section.

For application of the \emph{direct} estimation the data typically is
partitioned into a set of bins. Let us now also introduce a number of
$B$ non-overlapping bins, that cover the entire measurement
region. Each bin $i$ is associated with certain subset $\mathcal{B}_i$
of the measurement space implying that each measurement on the systems
can be assigned to exactly one bin. Hence we can introduce
non-overlapping sets of indexes
\begin{equation}
  \label{eq:index-sets}
  \mathcal{A}_i=\left\{j\in\left\{0,\ldots,N-1\right\}|x_j\in
    \mathcal{B}_i\right\}\, \mbox{for}\ i=1,...,B
\end{equation}
containing the indexes of all measurements within the respective
bins. In addition each bin $i$ is assigned a position $X_i$ by the
centre of mass of the respective set $\mathcal{B}_i$.

We now assume, that drift and diffusion are constant within the
individual bins, i.e. $\done(x)=\done_i$ and $\dtwo(x)=\dtwo_i\
\forall x\in\mathcal{B}_i$. These stepwise constant parts of the drift
and diffusion functions constitute the parameters we aim to optimize,
i.e.\ we have
$\boldsymbol{\Omega}=\left(\done_1,\ldots,\done_B,\dtwo_1,\ldots,\dtwo_B\right)$. If
the data is sampled at high frequency the small-$\tau$ approximation
of the propagator, (\ref{eq:propagator-small-tau}), can be used and
the log-likelihood function (\ref{eq:log_likelihood_function})
degenerates into contributions from the individuals bins,
\begin{subequations}
  \label{eq:loglikely-stepwise}
  \begin{eqnarray}
    \label{eq:loglikely-stepwise-long}
    L(\boldsymbol{\Omega})&=&\sum
    \limits_{i=1}^B\sum\limits_{j\in\mathcal{A}_i}\left[-\frac{\left(x_{j+1}-x_{j}-\done_i\tau\right)^2}{4\dtwo_i\tau}-\frac{1}{2}\log\left(4\pi\dtwo_i\tau\right)\right]\\
    \label{eq:loglikely-def-li}
    &=&\sum
    \limits_{i=1}^BL_i(\done_i,\dtwo_i)\quad.
  \end{eqnarray}
\end{subequations}
Since the summands $L_i$ depend on $\done_i$ and $\dtwo_i$ in the
respective bins only, their contributions are independent of one
another. Maximizing the joint likelihood function hence is equivalent
to maximizing the contributions of the individual bins. This
consequence of the finite-$\tau$ approximation is what makes the
maximum likelihood approach feasible for the non-parametric estimation
of stepwise-constant drift and diffusion functions.
\\
If we introduce frequencies $n_i:=\sum_{j\in\mathcal{A}_i}1$ and the
conditional moments
\begin{equation}
  \label{eq:conditional-moments}
  m^{\mathrm{(1)}}_i:=\frac{1}{n_i}{\sum_{j\in\mathcal{A}_i}\left(x_{j+1}-x_{j}\right)}\quad\mbox{and}\quad
  m^{\mathrm{(2)}}_i:=\frac{1}{n_i}{\sum_{j\in\mathcal{A}_i}\left(x_{j+1}-x_{j}\right)^2}
\end{equation}
the contributions of the individual bins are represented through
\begin{equation}
  \label{eq:loglikely-bin-i-efficient}
  L_i(\done_i,\dtwo_i)=-\frac{n_i}{2}\left[\frac{m^{\mathrm{(2)}}_i-2m^{\mathrm{(1)}}_i\done_i\tau+\left(\done_i\tau\right)^2}{2\dtwo_i\tau}+\log\left(4\pi\dtwo_i\tau\right)\right]
  \quad.
\end{equation}

For calculation of the log-likelihoods, hence, not the entire data set
is required. Instead the conditional moments $m^{\mathrm{(1)}}_i$ and
$m^{\mathrm{(2)}}_i$ and the frequencies $n_i$ are sufficient, which
need to be calculated only once. Subsequent evaluation of the
log-likelihood function then is computationally very efficient and
independent of the actual size of the data set. As in section
\ref{sec:estim-param-drift} maximisation of
(\ref{eq:loglikely-bin-i-efficient}) can be performed analytically and
yields
\begin{equation}
  \label{eq:nonparam-opt}
  \donetilde_i=\frac{1}{\tau}m^{\mathrm{(1)}}_i\quad\mbox{and}\quad
  \dtwotilde_i=\frac{1}{2\tau}\left(m^{\mathrm{(2)}}_i-\left[\tau\donetilde_i\right]^2\right)\quad.
\end{equation}

Apparently the maximum likelihood estimate for the drift term is equal
to the respective result from the \emph{direct} estimation procedure
(\ref{eq:direst-est}). The same holds for the diffusion term in the
limit $\tau\to 0$. At finite $\tau$ the diffusion estimate for bin $i$
obtained from the \emph{direct} estimation procedure differs from the
maximum likelihood estimate $\dtwotilde_i$ by
$\frac{1}{2}\tau(\donetilde_i)^2$. Interestingly this term has been
discussed as a potential finite-$\tau$ correction of the estimation
procedure \cite{Ragwitz01,FriedrichComment02,RagwitzComment02}.

\begin{figure*}
  \vspace*{-1cm}\includegraphics[width=7.5cm]{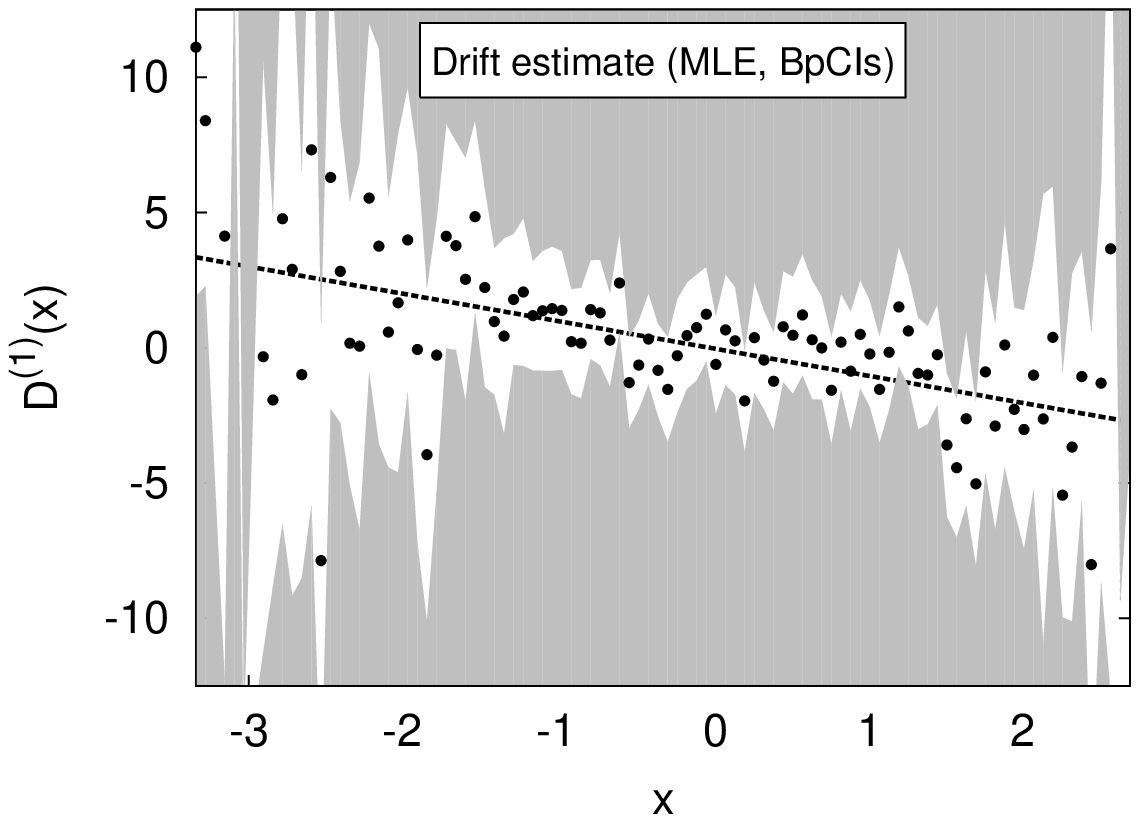}\hspace{.2cm}\includegraphics[width=7.5cm]{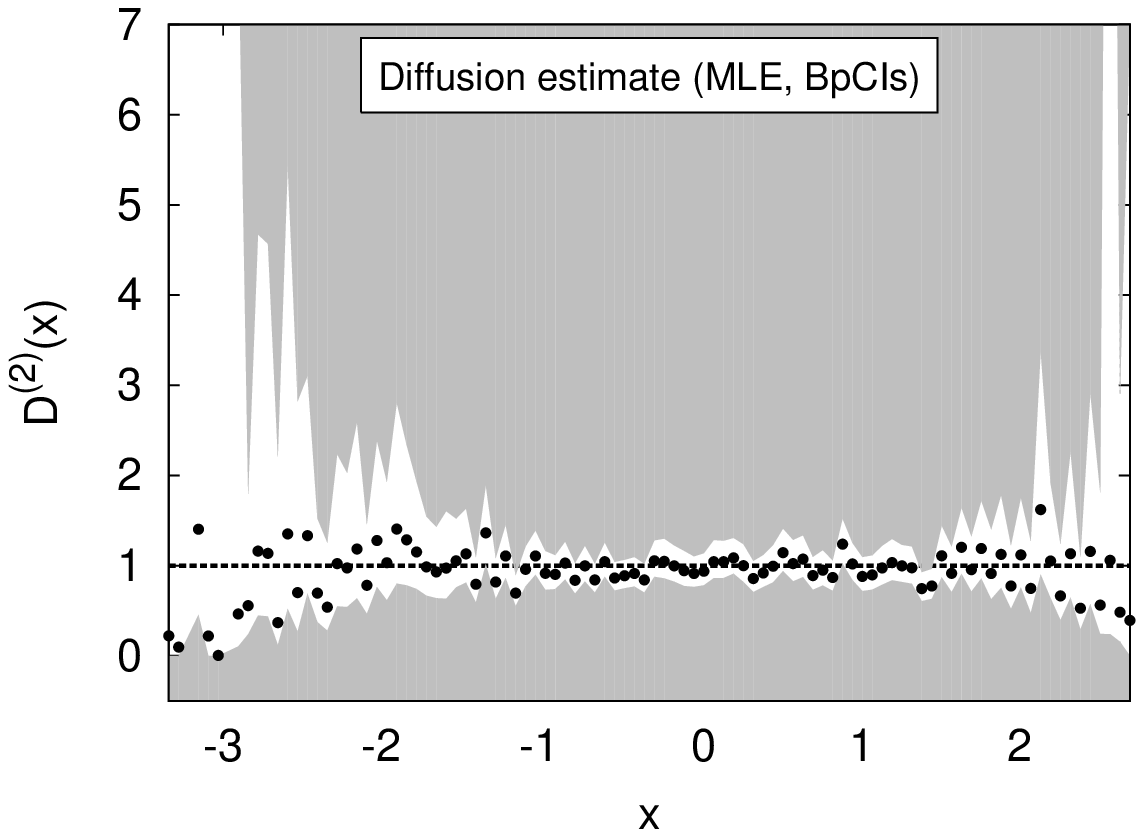}\\
  \includegraphics[width=7.5cm]{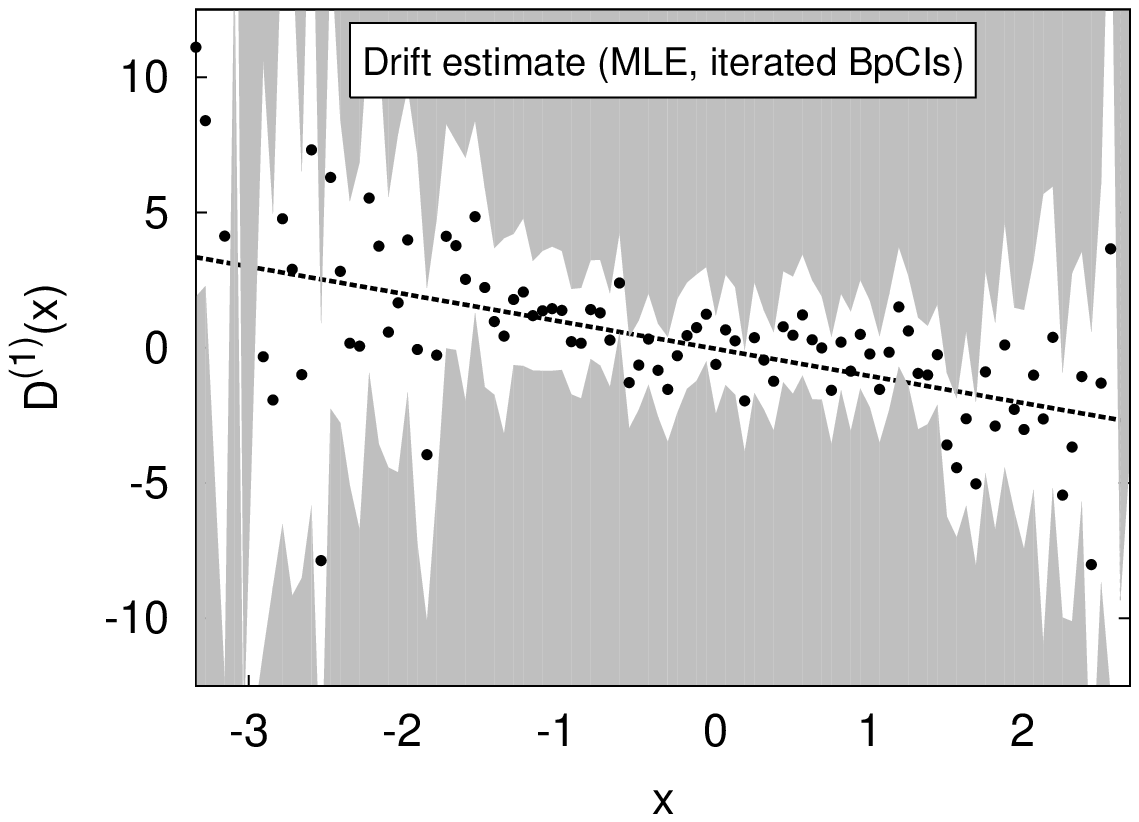}\hspace{.2cm}\includegraphics[width=7.5cm]{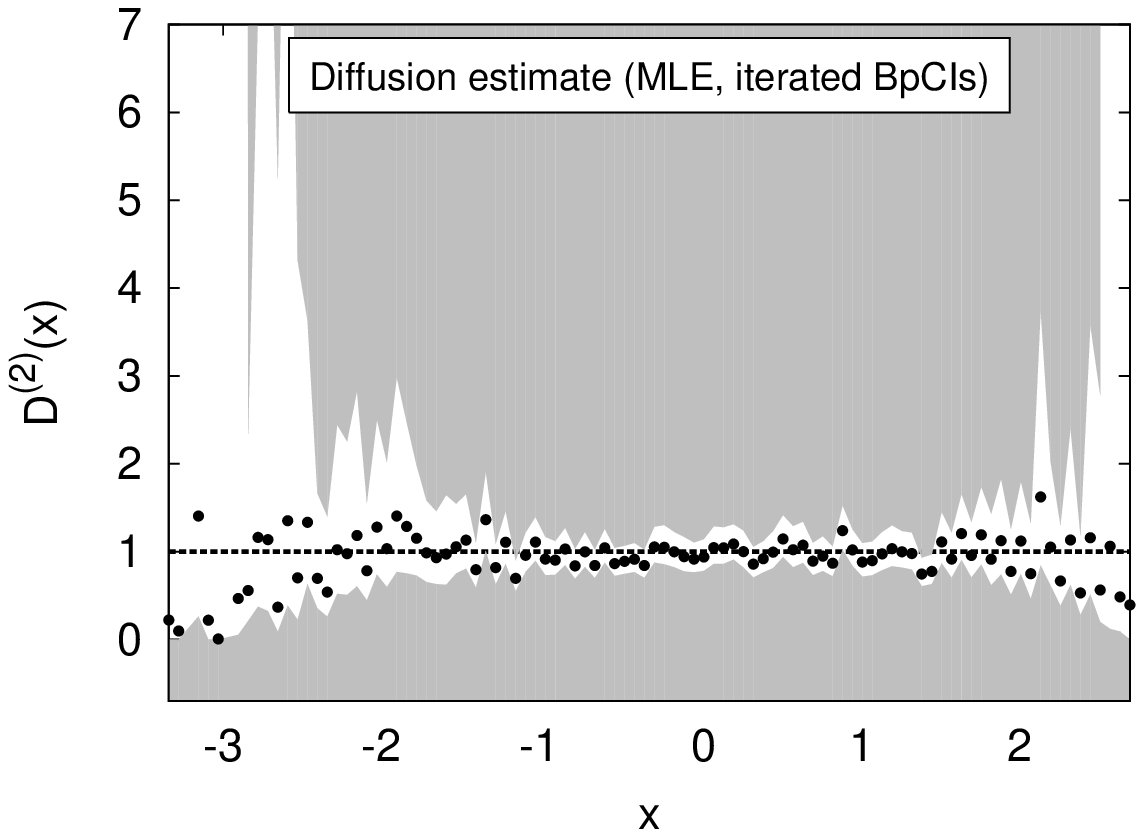}\\
  \includegraphics[width=7.5cm]{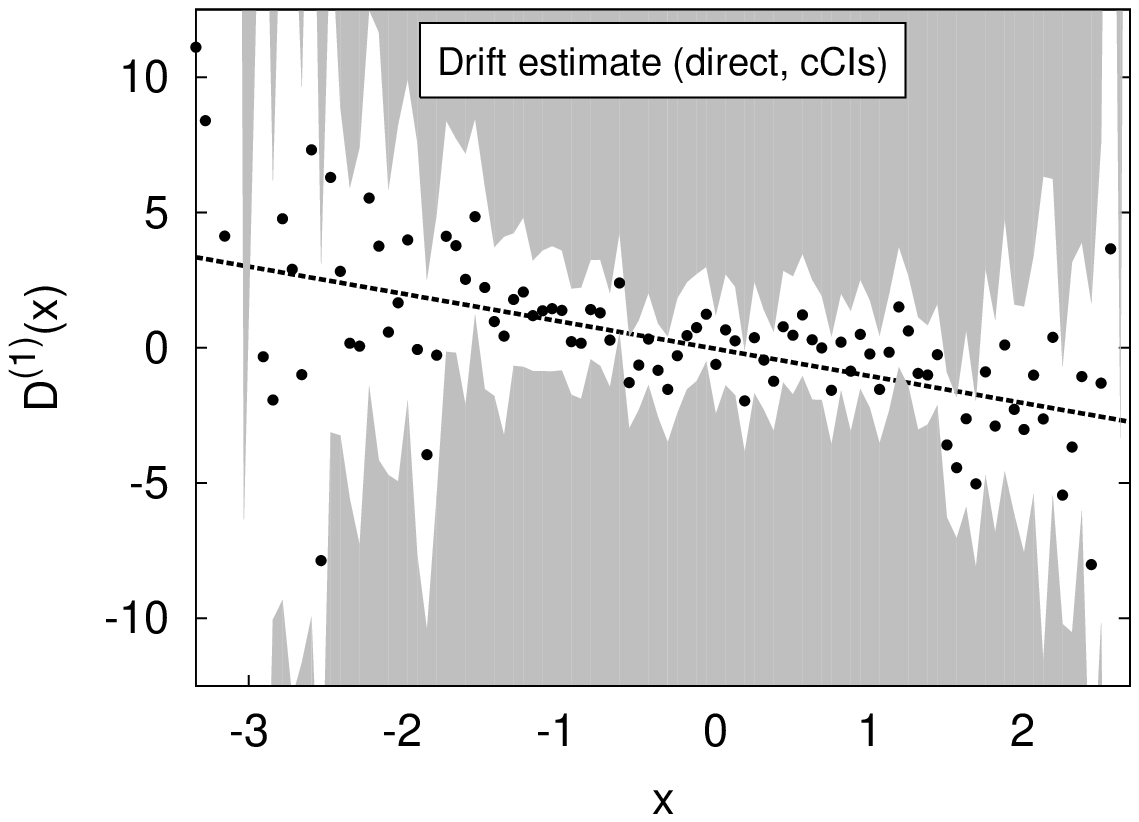}\hspace{.2cm}\includegraphics[width=7.5cm]{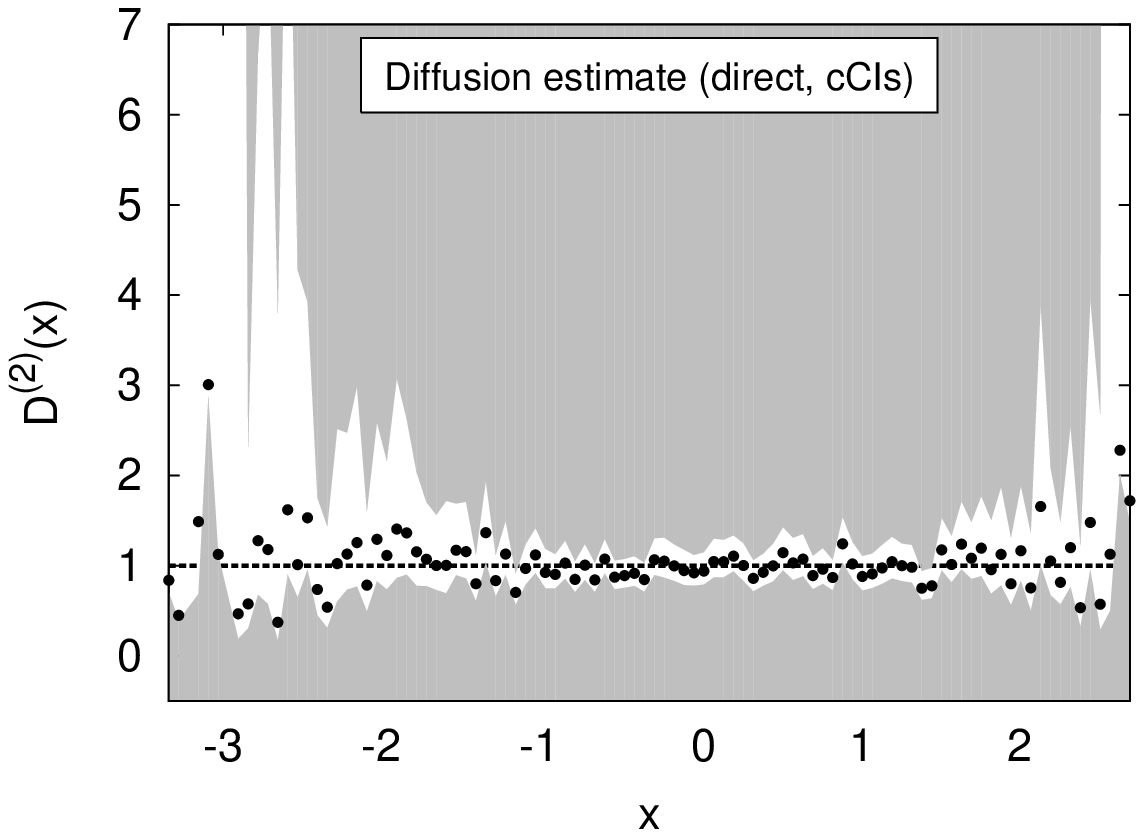}
  \caption{\label{fig:nonpar-estimates}Comparison of the
    non-parametric estimates for drift (left column) and diffusion
    (right column) functions obtained from maximum likelihood
    estimation as described in section \ref{sec:non-param-estim} (two
    upper rows) with the results from the \emph{direct} estimation
    (lowest row). All analyses were performed on a sample of length
    $N=10000$ of an Ornstein-Uhlenbeck process with $\gamma=1$ and
    $Q=1$ sampled at time increment $\tau=0.01$. The corresponding
    analytical drift and diffusion functions are indicated by the
    dashed lines. For analysis the data interval was partitioned into
    $B=100$ bins of equal size. The points indicate the respective
    estimates and the shades parameter regions outside the Bayesian
    posterior confidence intervals (for the maximum likelihood
    estimates, upper two rows) and the classical confidence interval
    (for the \emph{direct} estimates, lower row) as described in
    appendix \ref{sec:conf-interv-non}, each at $95\%$ level. For the
    uppermost row the offsets of the log-likelihoods were calculated
    from Wilks' theorem, whereas the offsets for the second row were
    iterated until the desired confidence was reached. Results of the
    iteration procedure are detailed in figure
    \ref{fig:nonpar-estimates-diff-sig}. Overall the procedures yield
    similar results, also with respect to the confidence intervals.
    Iteration of the log-likelihood offset (innermost row) results in
    increased Bayesian posterior confidence intervals for the
    diffusion estimates, in particular in sparsely sampled regions.}
\end{figure*}

\begin{figure}
  \includegraphics[width=7.5cm]{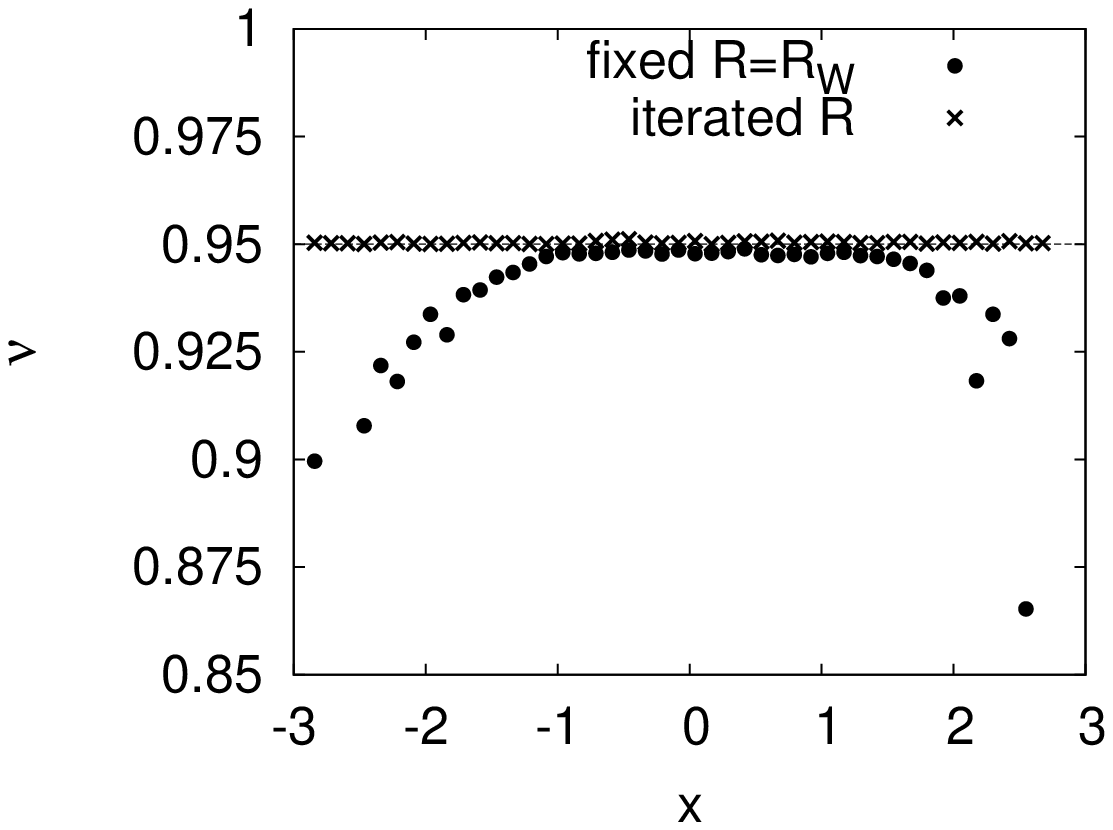}
  \caption{\label{fig:nonpar-estimates-diff-sig}Significance levels of
    Bayesian posterior confidence intervals for the non-parametric
    maximum likelihood estimates of the diffusion functions shown in
    figure \ref{fig:nonpar-estimates}. The data is described in detail
    in the caption of figure \ref{fig:nonpar-estimates}. The points
    indicate significance levels calculated from
    (\ref{eq:realised-confidence-binned-diff}) for confidence
    intervals determined numerically for fixed
    $R_i^{\mathrm{(2)}}=\rw(0.95)$ (red dots, corresponding to the
    upper row in figure \ref{fig:nonpar-estimates}) and for offsets
    $R_i^{\mathrm{(2)}}$ iterated individually for each bin with the
    aim to achieve the desired significance level $\nu=0.95$ (blue
    triangles, second row of figure
    \ref{fig:nonpar-estimates}). Obviously iteration of
    $R_i^{\mathrm{(2)}}$ in the individual bins in successful in
    achieving confidence intervals at a certain significance level
    (here: 0.95).}
\end{figure}

Beyond the calculation of the most likely parameters, likelihood
functions also are required for the calculation of BpCIs. Details on
how this procedure can be performed for the particular functions
(\ref{eq:loglikely-bin-i-efficient}) are outlined in appendix
\ref{sec:conf-interv-non}, together with remarks on the calculation of
the respective cCIs for the \emph{direct} estimation procedure. Both
the maximum likelihood and the \emph{direct} estimation procedures
were applied to one realisation of an Ornstein-Uhlenbeck process with
$N=10000$ sampled at $\tau=0.01$. The estimation results are depicted
in figure \ref{fig:nonpar-estimates}. For the non-parametric maximum
likelihood estimation (two upper rows) two different realisation of
BpCIs are shown, one with constant $R_i^{\mathrm{(2)}}=\rw(0.95)$
(upper row) and one where the individual $R_i^{\mathrm{(2)}}$ were
iterated to achieve a constant significance level of $\nu=0.95$
(second row).  Figure \ref{fig:nonpar-estimates-diff-sig} demonstrates
the effect of the iteration of $R_i^{\mathrm{(2)}}$ on the confidence
levels of BpCIs for the diffusion estimate.

Results of the \emph{direct} estimation procedure for the diffusion
differ slightly from the maximum likelihood estimates
(\ref{eq:nonparam-opt}). These differences, however, only seem to be
relevant at the sparsely sampled boundaries, where the term
$[\tau\donetilde_i]^2$ can become large. In the central regions with a
sufficient amount of samples per bin differences in the respective
confidence intervals are relatively small. Here BpCIs correspond very
well to cCIs, a fact already mentioned in section
\ref{sec:bayes-post-conf}. In the sparsely sampled regions cCIs of
$\done$ generally are larger than the respective BpCIs, which is a
result of the deviation of Student's t-distribution from
normality. For the diffusion estimates the respective confidence
intervals in the sparse regions differ quite significantly: The BpCIs
with constant significance level are largest and, therefore, probably
most conservative. Comparison between the uppermost and the second row
and inspection of figure \ref{fig:nonpar-estimates-diff-sig} suggests
that Wilks' theorem might not safely be applicable in sparsely sampled
regions. All confidence intervals almost everywhere contain the true
parameter values indicated by the dashed lines.
\\
Generally the differences between the maximum likelihood estimates and
the \emph{direct} estimation procedure are almost negligible and the
potentially small improvement in the confidence intervals for $\dtwo$
in many cases hardly may justify the application of this more complex
estimation procedure. The non-parametric estimation generally suffers
from huge uncertainties in sparsely sampled regions. For this reason a
quantitative interpretation of the estimates often is difficult.

If a quantitative characterisation of the system is required one
possibility is to fit parametrized drift and diffusion functions to
the non-parametric estimates. From inspection of figure
\ref{fig:nonpar-estimates} we can understand that this procedure might
not be straightforward, since one typically is faced with several
questions: which regions should be fitted? How should one account for
the confidence intervals, in particular for the asymmetric ones for
the diffusion estimate? In particular the latter question is not as
trivial as it seems.\\
If we for convenience assume that the confidence intervals for $\dtwo$
are symmetrical and that they correspond to quantiles of a normal
distribution standard procedures can be used to fit the
parametrizations (\ref{eq:ddfunc-5pars}) of drift and diffusion to the
\emph{direct} estimates for drift and diffusion exhibited in the lower
panels of figure \ref{fig:nonpar-estimates}. Two least square fits
were performed, one fitting the estimates only (NP-DE) and another
weighting the estimated by the reciprocal squared width of the
respective confidence intervals (NP-DE-CI). The results are compiled
in table \ref{tab:ou-small-tau-comparisson} and compared to the
results from maximum likelihood estimation of the parameters as
described in the previous section (MLE).  The accuracy of the
parameter estimates NP-DE and NP-DE-CI actually is fine, having in
mind that cCIs were not considered for the estimation (NP-DE) or
assumed to be symmetric (NP-DE-CI). Anyway the precision of the
estimates and the respective confidence intervals seems to be
deficient with respect to the estimation of parameters by maximum
likelihood methods.

\begin{turnpage}
  \begin{table}
    \begin{tabular}{l@{\hspace{1em}}rl@{\hspace{1em}}rl@{\hspace{1em}}rl@{\hspace{1em}}rl@{\hspace{1em}}rl@{\hspace{1em}}r}
      \hline\hline
      \multicolumn{1}{c}{Method}&\multicolumn{2}{c}{$\tilde{\Omega}_1$}&\multicolumn{2}{c}{$\tilde{\Omega}_2$}&\multicolumn{2}{c}{$\tilde{\Omega}_3$}&\multicolumn{2}{c}{$\tilde{\Omega}_4$}&\multicolumn{2}{c}{$\tilde{\Omega}_5$}&\multicolumn{1}{c}{Time}
      \\\hline
      MLE&$-0.98 $&$( -1.28, -0.69) $&$ 0.05$&$( -0.13, 0.24) $&$ -0.05$&$( -0.15, 0.03)$&$ 0.97$&$( 0.95, 1.00) $&$0.01 $&$( -0.01,0.03 )$&$364.7\,\mathrm{s}$\\
      NP-DE &$-0.58 $&$( -1.87, 0.71) $&$0.16$&$( -0.07, 0.39) $&$ -0.15$&$( -0.38, 0.08)$&$ 0.92$&$( 0.76, 1.08) $&$0.05 $&$(-0.01 ,0.09)$&$0.2\,\mathrm{s}$\\
      NP-DE-CI &$-0.69 $&$( -1.14, -0.24) $&$-0.05$&$( -0.23, 0.13) $&$ -0.19$&$( -0.29, -0.09)$&$ 1.02$&$( 0.99, 1.05) $&$-0.10 $&$(-0.10 ,-0.09)$&$0.2\,\mathrm{s}$\\
      BIN-MLE &$-1.01 $&$( -1.31, -0.72) $&$ 0.07$&$( -0.12, 0.26) $&$ -0.04$&$( -0.14, 0.05)$&$ 0.97$&$( 0.95, 1.00) $&$ 0.01$&$(-0.01 ,0.03 )$&$5.0\,\mathrm{s}$\\
      \hline\hline
    \end{tabular}
    \caption{\label{tab:ou-small-tau-comparisson}Parameters $\tilde{\Omega}_1$ to
      $\tilde{\Omega}_5$ of the drift and
      diffusion functions (\ref{eq:ddfunc-5pars}) estimated from realisations of an Ornstein-Uhlenbeck
      with $\Omega_1=-1$, $\Omega_4=1$, and $\Omega_2=\Omega_3=\Omega_5=0$.
      All estimates are based on the same data set of
      length $N=10000$
      sampled numerically at $\tau=0.01$. The estimates in the
      first row, MLE, are obtained from maximization of the
      likelihood function (\ref{eq:log_likelihood_function}) with the
      finite-$\tau$ approximation (\ref{eq:propagator-small-tau}). The
      second and third rows contain estimates obtained from fitting of
      the \emph{direct} estimates for drift and diffusion as plotted in
      the lower panels of figure \ref{fig:nonpar-estimates} with the
      respective polynomials. The fits for NP-DE rely only on the
      estimated values, whereas NP-DE-CI takes into account the
      respective  (classical) confidence intervals by weighting the
      estimates with the reciprocal of their squared width. The results in the last line, BIN-MLE, are obtained from
      maximization of the likelihood function (\ref{eq:bin-par-mle-loglikelihood-omega}) as outlined in section
      \ref{sec:effic-appr-estim}. The column \emph{Time} lists the
      computational time required for calculation of the respective estimates. Examples
      for the numerical implementation of the three estimation procedures
      used for generation of this benchmark are provided in appendix \ref{sec:pyth-code:-impl}.}
  \end{table}
\end{turnpage}

\section{\label{sec:effic-appr-estim} Estimation at small time
  increment and stepwise constant drift and diffusion: An
  efficient parametric procedure}
The estimation procedure for parametric drift and diffusion as
described in section \ref{sec:estim-gener-drift} yields accurate
estimation results and Bayesian posterior confidence regions if
measurement data is available at high sampling frequencies. A drawback
of this method, however, is that the calculation of the likelihood
functions becomes computationally demanding for large data sets or a
complex parameter space. This disadvantage can be solved by combining
it with the concept of stepwise constant drift and diffusion functions
developed in the previous section.

Let us assume that we have partitioned the data into $B$
non-overlapping bins, calculated the respective centres of mass $X_i$,
the index sets $\mathcal{A}_i$ from (\ref{eq:index-sets}), the
frequencies $n_i$, and the conditional moments according to
(\ref{eq:conditional-moments}). For a given combination of drift and
diffusion coefficients in the individual bins,
$\left(\done_1,\ldots,\done_B,\dtwo_1,\ldots,\dtwo_B\right)$, the
log-likelihood function then can be calculated as
\begin{equation}
  \label{eq:bin-par-mle-loglikelihood}
  L\left(\done_1,\ldots,\done_B,\dtwo_1,\ldots,\dtwo_B\right)=\sum_{i=1}^BL_i\left(\done_i,\dtwo_i\right)\quad,
\end{equation}
where the contributions of the individual bins are given by
(\ref{eq:loglikely-bin-i-efficient}). If we have a parametric form of
$\done$ and $\dtwo$ depending on a set of parameters
$\boldsymbol{\Omega}$ such as e.g.\ (\ref{eq:ddfunc-5pars}) the
coefficients in the individual bins for a particular
$\boldsymbol{\Omega}$ can be defined as
\begin{equation}
  \label{eq:coefficients-by-parametric-functions}
  \done_i:=
  \done\left(X_i,\boldsymbol{\Omega}\right)\quad\mbox{and}\quad   \dtwo_i:=
  \dtwo\left(X_i,\boldsymbol{\Omega}\right) \quad\forall i\quad,
\end{equation}
where the functions are evaluated at the centre of mass of the
respective bins. Then, equation (\ref{eq:bin-par-mle-loglikelihood})
can be used for calculation of the log-likelihood of
$\boldsymbol{\Omega}$,
\begin{equation}
  \label{eq:bin-par-mle-loglikelihood-omega}
  L\left(\boldsymbol{\Omega}\right)=\sum_{i=1}^BL_i\left(\done\left(X_i,\boldsymbol{\Omega}\right),\dtwo\left(X_i,\boldsymbol{\Omega}\right)\right)\quad.
\end{equation}
This expression needs to be maximized with respect to
$\boldsymbol{\Omega}$. An example for the numerical implementation
procedure is included in appendix \ref{sec:pyth-code:-impl} (procedure
\emph{parBinMLE}).  Results for optimizing the drift and diffusion
functions (\ref{eq:ddfunc-5pars}) for synthetic data obtained from
sampling an Ornstein-Uhlenbeck process are listed in table
\ref{tab:ou-small-tau-comparisson} (case BIN-MLE, last row).

Optimising expression (\ref{eq:bin-par-mle-loglikelihood-omega}) has
advantages with respect to the procedures for parametric estimation
outlined in sections \ref{sec:estim-gener-drift} and
\ref{sec:non-param-estim}: With respect to the optimisation of the
exact small-$\tau$ log-likelihood function,
(\ref{eq:log_likelihood_function}) with propagator
(\ref{eq:propagator-small-tau}), the computationally effort is reduced
drastically, in particular for large data sets. Once pre-calculation
of the conditional moments is completed the effort for evaluating the
log-likelihood does not depend on the dimension of the data set but of
the number of bins only. This is an important improvement, in
particular if the dimension $m$ of $\boldsymbol{\Omega}$ becomes large
and complex maximization algorithms are applied. The significant gain
in computational efficiency already has an effect at data sets
consisting of $N=10000$ points, where the computational effort roughly
is decreased by the order $10^2$ (rightmost column of table
\ref{tab:ou-small-tau-comparisson}).\\
With respect to the procedure outlined in the previous section, i.e.\
calculation of a non-parametric estimate and fitting the respective
functions, maximizing the log-likelihood yields much more precise
results (table \ref{tab:ou-small-tau-comparisson}). Moreover the
calculation of BpCIs is a consistent procedure, whereas fitting of
$\dtwo$ with its asymmetric confidence intervals cannot be performed
with many standard tools. Finally maximization of
(\ref{eq:coefficients-by-parametric-functions}) in contrast to
procedures for the non-parametric estimation is not very sensitive to
the binning, since (\ref{eq:bin-par-mle-loglikelihood-omega}) for bins
of equal size converges to (\ref{eq:log_likelihood_function}) with
propagator (\ref{eq:propagator-small-tau}) in the limit $B\to
\infty$. The estimation results listed in table
\ref{tab:ou-small-tau-comparisson} indicate that the procedure for the
sample data set already is sufficiently accurate with $B=100$.

\section{\label{sec:conclusions}Conclusions}
This work addresses the estimation of drift and diffusion functions
from time series data by means of maximum likelihood estimators with a
particular focus on the statistical accuracy of results through the
calculation of confidence intervals for the estimated parameters. The
results are discussed in the light of a procedure for the
\emph{direct} estimation of drift and diffusion function,
(\ref{eq:direst-est}), which extensively has been used in several
applications.

From the cases studied here we can conclude that maximum likelihood
estimators are superior to the direct estimation procedure if
quantitative results are required. However for a long time their
application for the estimation of general drift and diffusion
functions hardly was possible at reasonable effort, since it
intensively involved numerical solutions of the Fokker-Planck equation
(sections \ref{sec:estim-param-drift} and
\ref{sec:bayes-post-conf}). Within the scope of this work more
efficient methods for the estimation of parametric drift and diffusion
functions were developed.

In section \ref{sec:estim-gener-drift} an approximated log-likelihood
function for the estimation from data sampled at high frequencies is
introduced, which renders numerical solution of the Fokker-Planck
equation unnecessary. Although estimators in this limit already were
described earlier \cite{Sahalia02} their performance, at least to my
knowledge, was not investigated systematically. From our
investigations we see that the estimator yields good results even at
small but finite time lags (table \ref{tab:ou-small-tau-nobins}). The
numerical efficiency drastically can be increased if the data, in
addition, is partitioned into bins as described in section
\ref{sec:effic-appr-estim}. The application of this estimator e.g.\
for fitting drift and diffusion functions by low order polynomials
such as (\ref{eq:ddfunc-5pars}) is only slightly more complex than the
direct estimation procedure.  For this reason I would like to
recommend this procedure for the quantitative estimation of parametric
drift and diffusion functions from data available at sufficiently high
sampling frequency.

An advantage of the maximum likelihood approach is, that it is
straightforward to define Bayesian posterior confidence intervals
(BpCIs) for any parameter estimated. To my knowledge this problem had
not been addressed in connection with the estimation of drift and
diffusion functions, although of very high relevance for the
quantitative discussion of results. The calculation of BpCIs is
explained in great detail in section \ref{sec:bayes-post-conf} and
demonstrated by the respective examples in sections
\ref{sec:estim-param-drift} to \ref{sec:effic-appr-estim}.

In section \ref{sec:non-param-estim} the non-parametric estimation of
drift and diffusion by maximum-likelihood methods is addressed and
discussed in the context of the \emph{direct} estimation procedure,
(\ref{eq:direst-est}).  A similar procedure also is investigated in a
recent arXiv preprint by Ohkubo which uses kernel density estimators
for the transition probability density functions \cite{Ohkubo11}. In
contrast we apply a partition of the data into non-overlapping bins,
which is more straightforward to implement and which results in a
simple connection between global and local likelihood functions,
(\ref{eq:loglikely-stepwise}), a fact that is essential for the
efficient parametric estimation procedure developed subsequently.
\\
It is shown that the \emph{direct} estimation slightly differs from
the most likely estimates only in the diffusion term. Interestingly
these differences correspond to finite-$\tau$ correction for the
\emph{direct} estimation procedure proposed in \cite{Ragwitz01} that
was debated intensively \cite{FriedrichComment02,RagwitzComment02}.
Results derived in section \ref{sec:non-param-estim} and in appendix
\ref{sec:conf-interv-non} also are relevant for application of the
\emph{direct} estimation procedure, which until now lacked a detailed
discussion of the statistical errors. It is expected that the
resemblance to the maximum likelihood estimates and the assumption
required for this purpose (i.e.\ the small-$\tau$ approximation and
the assumption of stepwise constant drift and diffusion functions, see
section \ref{sec:non-param-estim} for details) also might ease the
interpretation of \emph{direct} estimates of drift and diffusion.

A great deal of the methods developed in this work relies on the
small-$\tau$ approximation of the propagator,
(\ref{eq:propagator-small-tau}), which only is applicable at
\emph{sufficiently} large sampling frequencies. It is difficult to
give an exact and general criterion which frequencies are required in
each case. As a rule of thumb the sampling, however, needs to allow
both for resolving the microscopic fluctuations at small time scales
and a sufficiently large fraction of phase space. For cases where the
accuracy of the small-$\tau$ approximation is questionable it is
recommended to validate the results by methods not explicitly
incorporating this approximation as e.g.\ described in section
\ref{sec:estim-param-drift} and in \cite{Kleinhans05,Kleinhans07MLE}.

At several instances the efficacy of the approximates for the
log-likelihood functions and the straightforward implementation of the
estimation procedure were emphasized. The computational costs of the
different approaches are compared in table
\ref{tab:ou-small-tau-comparisson} and show, that in particular the
computational effort for the procedure outlined in section
\ref{sec:effic-appr-estim} is very low, in spite of the optimization
of a five-dimensional parameter vector. The \emph{Python} codes used
for benchmarking including the implementation of the respective
estimation procedures are listed in appendix
\ref{sec:pyth-code:-impl}. I hope that these examples give ideas about
potential implementations of the estimation procedure and stimulate
the application of maximum likelihood estimators and Bayesian
posterior confidence intervals for the analysis of relevant
measurements.

\section*{Acknowledgements}%
I kindly would like to acknowledge discussions with Rudolf Friedrich,
Joachim Peinke, Maria Haase, Christoph Honisch, Bernd Lehle, and Pedro
Lind at the Dynamic Days Europe 2011, where this work was presented
first. Christoph Honisch furthermore pointed me to reference
\cite{Ohkubo11}, which covers a topic similar to section
\ref{sec:non-param-estim} of this work, although from a slightly
different perspective. I was glad to be able to include this
contribution in the discussion. All numerical simulations and
calculations were performed with \emph{Python} and I thank Frank
Raischel for introducing me to this language recently and Philip Rinn
for comments on the code. This work was supported by a Linnaeus-grant
from the Swedish Research Councils, VR and Formas
(http://www.cemeb.science.gu.se).

\section*{Appendix}
\appendix

\section{\label{sec:conf-interv-non}Confidence intervals for the
  non-parametric estimates}
This appendix addresses the calculation of Bayesian posterior
confidence intervals (BpCIs) for the non-parametric maximum likelihood
estimates and classical confidence intervals (cCIs) for the
\emph{direct} estimates, (\ref{eq:direst-est}). It uses nomenclature
and variables as introduced in section \ref{sec:non-param-estim}. The
respective confidence intervals were used for preparation of figure
\ref{fig:nonpar-estimates}.

For the individual bins BpCIs for drift and diffusion can be
calculated by the method introduced in section
\ref{sec:bayes-post-conf}. For the drift coefficients the expressions
can be evaluated analytically. With help of some algebra it can be
shown that the log-likelihood offset calculated from Wilks' theorem
independently of the sample size provides BpCIs exactly at the desired
confidence level.  For the drift in bin $i$ the BpCI at at level $\nu$
is
$\mathcal{C}_i^{\mathrm{(1)}}=\left[\donetilde_i-\sigma_i^{\mathrm{(1)}},\donetilde_i+\sigma_i^{\mathrm{(1)}}\right]$
with
\begin{equation}
  \label{eq:nonpar-confInt-drift}
  \sigma_i^{\mathrm{(1)}}=\sqrt{\frac{2}{n_i\tau}\dtwotilde_i}\sqrt{2}\,
  \mathrm{erf}^{-1}(\nu)\quad,
\end{equation}
where $\mathrm{erf}^{-1}(\nu)$ is the inverse of the error function at
the desired confidence level $\nu$. This interval is consistent with
classical confidence intervals for the mean of samples with known
variance (which is a consequence of the fact that we investigate
estimation errors in $\done_i$ and $\dtwo_i$ independent from one
another as described in section \ref{sec:bayes-post-conf}). For the
$95\%$ confidence level one obtains
$\sqrt{2}\,\mathrm{erf}^{-1}(0.95)=1.95996$.\\
As already previously noted for cCIs of direct estimates
\cite{Kleinhans05Euromech} the size of BpCIs for the drift estimates
diverges in the limit $\tau\to 0$, which renders the fact that the
estimation problem for the drift is ill-posed 
\cite{Pokern09}. Throughout this work we, however, could observe, that
estimation at small but finite $\tau$ yields satisfactory results and
that the respective confidence intervals accurately reflected the
statistical uncertainties of the drift estimates relevant at small
time lags.

BpCIs of the diffusion estimates to our knowledge cannot be derived
analytically but need to be iterated numerically from equation
(\ref{eq:likelihood-ratio-i}).  For $n_i>2$ at least a closed
expression of the confidence level $\nu_i^{\mathrm{(2)}}(R)$ of the
BpCI $\left[a_i(R),b_i(R)\right]$ associated to a certain $R$ can be
calculated by inserting (\ref{eq:loglikely-bin-i-efficient}) into
(\ref{eq:conf-level-nu-of-R-i}),
\begin{equation}
  \label{eq:realised-confidence-binned-diff}
   \nu^{\mathrm{(2)}}_i(R)=F_{\chi^2}\left(\frac{n_i}{a_i(R)}\dtwotilde_i,{n_i}-2\right)-F_{\chi^2}\left(\frac{n_i}{b_i(R)}\dtwotilde_i,
    {n_i}-2\right)\quad,
\end{equation}
where $F_{\chi^2}(x,k)$ is the cumulative distribution function of the
$\chi^2$ distribution at $x$ with $k$ degrees of freedom.  This
expression can either be used for verification of the confidence level
for a constant $R$ or for iteration of $R$ (and boundaries of the
corresponding confidence interval) until a desired confidence level is
reached.  For bins $i$ with $n_i<3$ BpCIs for the diffusion estimate
cannot be calculated since (\ref{eq:conf-level-nu-of-R-i}) diverges.

For the direct estimates, (\ref{eq:direst-est}), classical confidence
intervals (cCIs) at confidence level $\nu$ can be calculated from
quantiles of  Student's t and the  $\chi^2$
distribution, respectively. For the drift estimates the confidence
intervals are symmetric with respect to the estimated drift
coefficients. For bin $i$ the cCI at level $\nu$ is
$\left[\donehat_i-\sigma_i^{\mathrm{(1)}},\donehat_i+\sigma_i^{\mathrm{(1)}}\right]$
with
\begin{equation}
  \label{eq:direct-confidence-drift}
  \sigma_i^{\mathrm{(1)}}:=\sqrt{\frac{2}{n_i\tau}\dtwohat_i-\frac{1}{n_i}\left(\donehat_i\right)^2}F_t^{-1}\left(\frac{1+\nu}{2},n_i-1\right)\quad,
\end{equation}
where $F_t^{-1}(x,i)$ is the inverse cumulative distribution function
of Student's t-distribution with $i$ degrees of freedom, evaluated at
$x$. For the significance level $\nu=0.95$ relevant in this manuscript
for large sample sizes $F_t^{-1}(0.975,i\to\infty)=1.95996$ is
obtained, which equals the respective value for the BpCIs for the
drift estimates calculated above.

cCIs of the diffusion estimates are asymmetric. The boundaries
$\left[a_i,b_i\right]$ of the cCIs for the diffusion estimate in bin
$i$ at level $\nu$ can be calculated as
\begin{subequations}
  \label{eq:direct-confidence-diff}
  \begin{eqnarray}
    a_i&=&\frac{1}{2}\tau\left(\donehat_i\right)^2+n_i\frac{\dtwohat_i-\frac{1}{2}\tau\left(\donehat_i\right)^2}{F_{\chi^2}^{-1}\left(\frac{1+\nu}{2},n_i-1\right)}\quad\mbox{and}\\
    b_i&=&\frac{1}{2}\tau\left(\donehat_i\right)^2+n_i\frac{\dtwohat_i-\frac{1}{2}\tau\left(\donehat_i\right)^2}{F_{\chi^2}^{-1}\left(\frac{1-\nu}{2},n_i-1\right)}\quad,
  \end{eqnarray}
\end{subequations}
where $F_{\chi^2}^{-1}(x,i)$ is the inverse cumulative distribution
function of the $\chi^2$ distribution with $i$ degrees of freedom
evaluated at $x$. Quantiles of distribution functions such as
Student's t or the $\chi^2$ distributions in the meanwhile accurately
are 
iterated numerically by standard packages as e.g.\ applied in the
\emph{Python} codes used for preparation of this manuscript as listed
in appendix \ref{sec:pyth-code:-impl}.

\section{Python code: Implementation of estimation procedures and
  benchmark\label{sec:pyth-code:-impl}}
This appendix contains listings of \emph{Python} code used for
preparation of the tables and figures contained in this
manuscript. \emph{Python} is available without charge for most software
environments. The codes were tested with version 2.6.5.

The file \emph{python\_procedures.py} contains examples for
implementations of the parametric estimation procedures described in
sections \ref{sec:estim-gener-drift} (\emph{parMLE}) and
\ref{sec:effic-appr-estim} (\emph{parBinMLE}). In addition an
implementation of the \emph{direct} estimation procedure for the
non-parametric estimation is included (\emph{nonParDirect}), that
yields classical confidence intervals for the estimates calculated
according to appendix \ref{sec:conf-interv-non}.

The file \emph{python\_main.py} exhibits an example of how the
individual procedures can be used on a specific data set. This code
has been used for calculation of the results and benchmarks listed in
table \ref{tab:ou-small-tau-comparisson}.

\definecolor{grey}{rgb}{.95,.95,.95}
\lstset{language=Python,basicstyle=\tiny,backgroundcolor=\color{grey},breaklines=false,title=\lstname}

\lstset{language=Python,basicstyle=\tiny,backgroundcolor=\color{grey},breaklines=false,title={MLE\_estimation\_procedures.py}}
\begin{lstlisting}
import scipy
import scipy.optimize
import scipy.stats
import scipy.special
import numpy

#===============================================================================
# parMLE - PARAMETRIC MLE WITH FINITE TIME APPROXIMATION               (SECT. 4)
#===============================================================================

def parNegLogLikelihood(pars,dat,deltat,d1,d2):
    #Log-likelihood function, dependent on functions d1, d2 and data 
    res=[]
    for i in range(len(dat)-1):
        var=2.*d2(pars,dat[i])*deltat
        if var>0:
            res.append(scipy.log(2.*scipy.pi*var)/2+
                       (dat[i+1]-(dat[i]+d1(pars,dat[i])*deltat))**2/(2.*var))
        else:
            res.append(float("Inf"))
    return sum(res)

def parConfFkt(conf,i,sigRatio,dat,deltat,d1,d2,paropt,opt):
    #Roots at boundaries of the 1d confidence interval of parameter i
    dev=numpy.zeros(len(paropt))
    dev[i]=conf
    return (opt-parNegLogLikelihood(paropt+dev,dat,deltat,d1,d2)
            -scipy.log(sigRatio))

def parMLE(dat,deltat,d1,d2,par0,desiredSig,xtol):
    res=numpy.zeros([len(par0),3])
    #sig=numpy.zeros(len(par0))
    sigRatio=scipy.exp(-scipy.stats.chi2.ppf(desiredSig,1)/2)

    #Find optimal parameters with simplex gradient method
    print "parMLE: Finding best parameters:"
    res[:,0]= scipy.optimize.fmin(parNegLogLikelihood,par0,
                                   args=[dat,deltat,d1,d2],xtol=xtol,disp=False)
    opt=parNegLogLikelihood(res[:,0],dat,deltat,d1,d2)

    #Calculate width of condidence intervals
    for i in range(len(par0)):
        print "Parameter ",i+1,":",res[i,0],", BPCI: (",
        #find lower bound for bisection root finding algorithm
        res[i,1]=-1.
        while parConfFkt(res[i,1],i,sigRatio,dat,deltat,d1,d2,res[:,0],opt)>=0:
            res[i,1]=res[i,1]-1.
        #bisection
        res[i,1]=scipy.optimize.bisect(parConfFkt,res[i,1],0.
                     ,args=(i,sigRatio,dat,deltat,d1,d2,res[:,0],opt),xtol=1e-3)

        #find upper bound for bisection root finding algorithm
        res[i,2]=1.
        while parConfFkt(res[i,2],i,sigRatio,dat,deltat,d1,d2,res[:,0],opt)>0:
            res[i,2]=res[i,2]+1.
        #bisection
        res[i,2]=scipy.optimize.bisect(parConfFkt,0.,res[i,2]
                     ,args=(i,sigRatio,dat,deltat,d1,d2,res[:,0],opt),xtol=1e-3)   
        print res[i,0]+res[i,1],',',res[i,0]+res[i,2],')'

#===============================================================================
# parBinMLE - PARAMETRIC MLE, FINITE TIME APPR. AND SPATIAL DISCRET.   (SECT. 6)
#===============================================================================

def binNegLogLikelihood(pars,incmean,incmeansq,n,deltat):
    #Log-likelihood function for bin-wise estimation based on statistical pars
    if (pars[1]>0):
        var=2.*pars[1]*deltat
        res=n/2*(scipy.log(2.*scipy.pi*var)
             +(incmeansq-2.*pars[0]*deltat*incmean+(pars[0])**2*deltat**2)/var)
    else:
        res=float("Inf")
    return res

def parBinNegLogLikelihood(pars,incmean,incmeansq,incn,xCenter,d1,d2,deltat):
    #Roots at boundaries of the 1d confidence interval of parameter i
    res=[]
    for i in range(len(incn)):
        res.append(binNegLogLikelihood([d1(pars,xCenter[i]),d2(pars,
                   xCenter[i])],incmean[i],incmeansq[i],incn[i],deltat))
    return sum(res)

def parBinConfFkt(conf,i,sigRatio,incmean,incmeansq,incn,xCenter,deltat,
                  d1,d2,paropt,opt):
    #Roots at boundaries of the 1d confidence interval of parameter i
    dev=numpy.zeros(len(paropt))
    dev[i]=conf
    return (opt-parBinNegLogLikelihood(paropt+dev,incmean,incmeansq,incn,
                                      xCenter,d1,d2,deltat)-scipy.log(sigRatio))

def parBinMLE(dat,deltat,nbin,d1,d2,par0,desiredSig,xtol):
    res=numpy.zeros([len(par0),3])
    sigRatio=scipy.exp(-scipy.stats.chi2.ppf(desiredSig,1)/2)

    #init binning
    xmax=numpy.max(dat)
    xmin=numpy.min(dat)
    deltax=(xmax-xmin)/nbin
    xLeft=numpy.array(range(nbin))*deltax+xmin
    xCenter=xLeft+deltax/2
    incmean=numpy.zeros(nbin)
    incmeansq=numpy.zeros(nbin)
    incn=numpy.zeros(nbin,dtype=numpy.int)

    #calculate statistics of bins
    for j in range(0,nbin):
        #select relevant data
        jDat=numpy.where((dat[:len(dat)-1]
                                   <xLeft[j]+deltax)&(dat[:len(dat)-1]>=xLeft[j]))
        incn[j]=len(jDat[0])
        inc=dat[jDat[0]+1]-dat[jDat[0]]
        incmean[j]=sum(inc)/incn[j]
        incmeansq[j]=sum(inc**2)/incn[j]

    #Find optimal parameters with simplex gradient method
    print "parBinMLE: Finding best parameters:"
    res[:,0]= scipy.optimize.fmin(parBinNegLogLikelihood,par0,
        args=[incmean,incmeansq,incn,xCenter,d1,d2,deltat],xtol=xtol,disp=False)
    opt=parBinNegLogLikelihood(res[:,0],incmean,incmeansq,incn,
                                                           xCenter,d1,d2,deltat)

    #Calculate width of condidence intervals
    for i in range(len(par0)):
        print "Parameter ",i+1,":",res[i,0],", BPCI: (",
        #find lower bound for bisection root finding algorithm
        res[i,1]=-1.
        while parBinConfFkt(res[i,1],i,sigRatio,incmean,incmeansq,
                                     incn,xCenter,deltat,d1,d2,res[:,0],opt)>=0:
            res[i,1]=res[i,1]-1.
        #bisection
        res[i,1]=scipy.optimize.bisect(parBinConfFkt,res[i,1],0.,
                              args=(i,sigRatio,incmean,incmeansq,
                              incn,xCenter,deltat,d1,d2,res[:,0],opt),xtol=1e-3)

        #find upper bound for bisection root finding algorithm
        res[i,2]=1.
        while parBinConfFkt(res[i,2],i,sigRatio,incmean,incmeansq,
                                      incn,xCenter,deltat,d1,d2,res[:,0],opt)>0:
            res[i,2]=res[i,2]+1.
        #bisection
        res[i,2]=scipy.optimize.bisect(parBinConfFkt,0.,res[i,2],
                              args=(i,sigRatio,incmean,incmeansq,
                              incn,xCenter,deltat,d1,d2,res[:,0],opt),xtol=1e-3)   
        print res[i,0]+res[i,1],',',res[i,0]+res[i,2],')'

#===============================================================================
# nonParDirect - DIRECT EST. (FINITE TIME APP. AND SPATIAL DISCRET) (APPENDIX A) 
#===============================================================================

def nonParDirect(dat,deltat,nbin,significanceLevel):
    #Direct estimation of drift and diffusion in nbin bins from time series

    #initialisation
    res=numpy.zeros([nbin,2,3])
    incn=numpy.zeros(nbin,dtype=numpy.int)

    #binning
    xmax=numpy.max(dat)
    xmin=numpy.min(dat)
    deltax=(xmax-xmin)/nbin
    xLeft=numpy.array(range(nbin))*deltax+xmin
    xCenter=xLeft+deltax/2

    #process individual bins sequentially
    print 'parBinMLE: Processing ',nbin,' bins sequentially:'
    for j in range(0,nbin):
        #select relevant data
        jDat=numpy.where((dat[:len(dat)-1]
                                   <xLeft[j]+deltax)&(dat[:len(dat)-1]>=xLeft[j]))
        incn[j]=len(jDat[0])

        if incn[j]>1:
            print j,xCenter[j],' ',

            #calculate increments
            inc=dat[jDat[0]+1]-dat[jDat[0]]
            incmean=sum(inc)/incn[j]
            incmeansq=sum(inc**2)/incn[j]
            incmeanquad=sum(inc**4)/incn[j]

            #Direct estimation of d1 and d2
            res[j,0,0]=incmean/deltat
            res[j,1,0]=incmeansq/deltat/2

            #Standard errors for d1
            res[j,0,2]=(scipy.stats.t.ppf((1.+significanceLevel)/2,incn[j]-1)
                        *scipy.sqrt((incmeansq-incmean**2)/incn[j])/deltat)
            res[j,0,1]=-res[j,0,2]

            #Standard errors for d2
            res[j,1,1]=deltat/2.*res[j,0,0]**2-res[j,1,0]+1.*incn[j]*(
              res[j,1,0]-deltat/2.*res[j,0,0]**2)/(
              scipy.stats.chi2.ppf((1.+significanceLevel)/2,1.*(incn[j]-1)))
            res[j,1,2]=deltat/2.*res[j,0,0]**2-res[j,1,0]+1.*incn[j]*(
              res[j,1,0]-deltat/2.*res[j,0,0]**2)/(
              scipy.stats.chi2.ppf((1.-significanceLevel)/2,1.*(incn[j]-1)))

            print 'D1: ',res[j,0,0],', CI: (',
            print res[j,0,0]+res[j,0,1],',',res[j,0,0]+res[j,0,1],') ',
            print 'D2: ',res[j,1,0],', CI: (',
            print res[j,1,0]+res[j,1,1],',',res[j,1,0]+res[j,1,2],') '
\end{lstlisting}

\lstset{language=Python,basicstyle=\tiny,backgroundcolor=\color{grey},breaklines=false,title={main.py}}
\begin{lstlisting}
#===  PARAMETERS FOR SIMULATION AND ANALYSIS  ==================================
#
#--  SIMULATION OF OU PROCESS
gamma=1.0
d=1.0
deltat=0.01
dim=10000
#
#-- MLE ESTIMATION
startingParameters=[1.,1.,1.,1.,1.]
desiredSig=0.95
#
#-- PARAMETRISED FUNCTIONS
#- DRIFT:
def d1(pars,x):
    return pars[0]*x+pars[1]*x**2+pars[2]*x**3
#
#- DIFFUSION:
def d2(pars,x):
    return pars[3]+pars[4]*x**2
#
#===============================================================================

import random
import scipy
import scipy.optimize
import numpy
import time
from MLE_estimation_procedures import *

#A) ======  SIMULATION OF DATA SET  ============================================
print "Generating OU-data ..."

x=numpy.array(numpy.zeros(dim))
sfact=scipy.exp(-gamma*deltat)
sdev=scipy.sqrt(d/gamma*(1.-scipy.exp(-2.*gamma*deltat)))

# SAMPLING OF THE FINITE TIME PROPAGATOR
for i in range(1,dim):
    x[i]=random.normalvariate(x[i-1]*sfact,sdev)

#B) ======  DATA ANALYSIS  =====================================================
print "Analysing data ..."
t=time.time()
 
# PARAMETRIC MLE WITH FINITE TIME APPROXIMATION
parMLE(x,deltat,d1,d2,startingParameters,desiredSig,xtol=1e-8)
print "Time elapsed: ",time.time()-t,"s"
t=time.time()

# DIRECT ESTIMATION PROCEDURE ACCORDING TO SIEGERT, FRIEDRICH, PEINKE ET AL.
nonParDirect(x,deltat,100,desiredSig)
print "Time elapsed: ",time.time()-t,"s"
t=time.time()

# PARAMETRIC MLE WITH FINITE TIME APPROXIMATION AND SPATIAL DISCRETIZATION(BINS)
parBinMLE(x,deltat,100,d1,d2,startingParameters,desiredSig,xtol=1e-8)
print "Time elapsed: ",time.time()-t,"s"
t=time.time()

print "... done."
\end{lstlisting}

\section*{References}

%
\end{document}